\documentclass[conference]{IEEEtran}
\IEEEoverridecommandlockouts

\usepackage{tikz}

\newcommand\copyrighttext{%
  \footnotesize \textcopyright 2021 IEEE. Personal use of this material is permitted. Permission from IEEE must be obtained for all other uses, in any current or future media, including reprinting/republishing this material for advertising or promotional purposes, creating new collective works, for resale or redistribution to servers or lists, or reuse of any copyrighted component of this work in other works.}
\newcommand\copyrightnotice{%
\begin{tikzpicture}[remember picture,overlay]
\node[anchor=south,yshift=10pt] at (current page.south) {\fbox{\parbox{\dimexpr\textwidth-\fboxsep-\fboxrule\relax}{\copyrighttext}}};
\end{tikzpicture}%
}

\newcommand\acceptedtext{%
  \footnotesize Accepted for the proceedings of Military Communications Conference 2021 - NDN Session (MILCOM 2021) [Camera-ready version]}
\newcommand\acceptednotice{%
\begin{tikzpicture}[remember picture,overlay]
\node[anchor=north,yshift=-30pt] at (current page.north) {\fbox{\parbox{\dimexpr\textwidth-\fboxsep-\fboxrule\relax}{\acceptedtext}}};
\end{tikzpicture}%
}

\usepackage{cite}
\usepackage{amsmath,amssymb,amsfonts}
\usepackage{algorithmic}
\usepackage{graphicx}
\usepackage{textcomp}
\usepackage{xcolor}
\def\BibTeX{{\rm B\kern-.05em{\sc i\kern-.025em b}\kern-.08em
    T\kern-.1667em\lower.7ex\hbox{E}\kern-.125emX}}

\graphicspath{{figure/}{figures/}}

\ifCLASSOPTIONcompsoc
	\usepackage[caption=false,font=normalsize,labelfont=sf,textfont=sf]{subfig}
\else
	\usepackage[caption=false,font=footnotesize]{subfig}
\fi

\begin{document}

\title{BLEnD: Improving NDN Performance Over Wireless Links Using Interest Bundling}

\author{\IEEEauthorblockN{Md Ashiqur Rahman}
\IEEEauthorblockA{
	\textit{The University of Arizona}\\
	marahman@cs.arizona.edu}
\and
\IEEEauthorblockN{Teng Liang}
\IEEEauthorblockA{
	\textit{Peng Cheng Laboratory}\\
	liangt@pcl.ac.cn}
\and
\IEEEauthorblockN{Beichuan Zhang}
\IEEEauthorblockA{
	\textit{The University of Arizona}\\
	bzhang@cs.arizona.edu}
}


\maketitle

\copyrightnotice
\acceptednotice

\vspace{-10pt}

\begin{abstract}
	Named Data Networking (NDN) employs small-sized Interest packets to retrieve large-sized Data packets. Given the half-duplex nature of wireless links, Interest packets frequently contend for the channel with Data packets, leading to throughput degradation. In this work, we present a novel idea called BLEnD, an Interest-bundling technique that encodes multiple Interests into one at the sender and decodes at the receiver. The major design challenges are to reduce the number of Interest transmissions without impacting the one-Interest one-Data principle embedded everywhere in NDN architecture and implementation, and support flow/congestion control mechanisms that usually use Interest packets as signals. BLEnD achieves these by bundling/unbundling Interests at the link adaptation layer, keeping all NDN components unaware and unaffected. Over a one-hop Wi-Fi link, BLEnD improves application throughput by 30\%. It may also be used over multiple hops and be improved in a number of ways.
\end{abstract}


\section{Introduction}
\label{sec:introduction}


Wireless communication is ubiquitous in tactical networks. Despite the increasing 
demand for high throughput and reliability for mission-critical applications, 
performance over wireless links is severely hampered, irrespective of single or multiple 
hops \cite{goldsmith2005wireless}. One significant contributing factor is the shared 
nature of a wireless channel where Data and signal packets cannot be exchanged 
simultaneously. In TCP-IP, the acknowledgment (ACK) is considered the signal packet 
to indicate a Data packet is received. In Named Data Networking (NDN)~\cite{ndn-ccr}, 
Interest packets act as the signals to fetch Data packets. Although signals are usually 
much smaller than data packets, they have similar medium access overhead (IEEE 802.11 
MAC protocol~\cite{ieee-mac}). Reducing the number of signal packets should make the 
channel more available to data packets, thus increasing overall goodput. The downside 
is the reduced feedback to the data sender for purposes such as congestion control and 
loss recovery. A well-designed signal packet reduction technique should offer a better 
tradeoff that increases throughput while supporting loss recovery and congestion control. 
These are crucial for challenging environments such as battlefields, be in an ad-hoc 
network of soldiers and vehicles or where troops are connected to an access point.

Existing works in wireless local area networks (WLANs) using TCP-IP show that reduced 
ACK frequency for multiple data packets reduces channel occupancy, hence, improves 
application throughput or goodput~\cite{tack-Tong, altman2003novel, floyd-profile, 
landstrom2007, nadh2013improving}. However, most of the works are based on the 
end-to-end IP architecture, directly implemented at the two endpoints.

NDN promises a more scalable architecture than IP through its name-based hop-by-hop 
design, especially for multi-hop ad-hoc networks~\cite{Rahman-daf-manet}. One of its 
core design choices, \textit{one Interest for one Data}, is embedded everywhere in 
the architecture, from applications through the link adaptation layer. It poses a 
significant challenge to the idea of reducing the frequency of signal packets, i.e., 
Interests, because that would cause the imbalance of Interest and Data packets.

Our initial test simulations over a single hop Wi-Fi showed that application goodput can 
be improved by up to 59.1\% for IEEE 802.11b  and 91.7\% for 802.11n, respectively, 
if only one Interest is used to retrieve an entire file (Fig.~\ref{fig:potential-improvement},
details in Sec.~\ref{subsec:interest-reduction}). Even so, trying to reduce the 
number of Interests to bring multiple Data packets will surely break NDN's stateful 
forwarding~\cite{Yi-stateful}. Thus, existing efforts in this direction (\cite{Ding-video, 
Fan-infocom-video}) require fundamental changes to NDN's packet forwarding mechanism 
with unknown impacts to other parts of the architecture.

Potential improvements by reducing Interest packets and the need to avoid fundamental 
NDN design changes motivated us to formulate a novel technique named BLEnD (\textbf{B}undling
Interests with \textbf{L}ink-layer \textbf{En}coding and \textbf{D}ecoding). 
It is a link-layer solution that \textit{hides} any encoding/decoding information 
from the upper layers. At the sender, BLEnD encodes multiple Interest packets into one before forwarding to the network interface card (NIC). It also includes information 
for the receiver to decode one bundled Interest into multiple individual ones before 
giving them to the local NDN forwarder. Thus BLEnD reduces the number of Interest 
transmissions over a wireless channel while maintaining the original number of Interests at the upper layers, making NDN components unaware and unaffected by the Interest 
bundling. 

To the best of our knowledge, this is the first work that considers channel utilization 
optimization in NDN directly at the link-layer without any change to the architecture and 
the latest packet format specifications~\cite{NDN-packet}. BLEnD currently achieves roughly 
32.57\% and 37.03\% better goodput over no bundling in 802.11b and 802.11n, respectively. 
These results are for a single-hop, infrastructure mode Wi-Fi communication. Such 
improvements also inspired us to conduct further analysis of challenges to formulate 
potential future solutions and explore application domains such as multi-hop ad-hoc networks.

%
\section{Motivation and Background}
\label{sec:motivation-background}

\subsection{Reducing Interest packet frequency increases goodput}
\label{subsec:interest-reduction}
Our primary motivation comes from the recent work by \cite{tack-Tong} in TCP-IP. Their design sends periodic ACK, which significantly improves the Data throughput. We followed their UDP baseline testing for our preliminary simulation analysis using ndnSIM~\cite{mastorakis2017ndnsim}. We simulated single-hop Wi-Fi communication in infrastructure mode using IEEE 802.11b (at 11 Mbps) and 802.11n (at 24 Mbps), with one consumer as the station node and the producer as the access point. Each Data packet payload size is 1460 bytes. First, we used the existing AIMD and CUBIC congestion controllers from the simulator with default Interest-Data pipelining in the NDN Forwarding Daemon (NFD)~\cite{afanasyev2014nfd} to collect a baseline lower bound. We then implemented a dummy application where the consumer sends only one Interest packet. Upon receiving, the producer continuously sends the data packets back. Fig.~\ref{fig:potential-improvement} shows that reducing Interest packets significantly improves goodput and gives us an estimated upper bound for BLEnD.

\begin{figure}[!t]
  \vspace{1mm}
  \scriptsize
  \centering
  \begin{tabular}{l|c|c}
    \hline
      & 802.11b & 802.11n \\
      & at 11 Mbps & at 24 Mbps \\
    \hline
    Default or no bundle (Mbps) & 4.4 &  9.7 \\
    \hline
    One Interest (Mbps) & 7.0 & 18.6 \\
    \hline
    Goodput improvement & 59.1\% & 91.7\% \\
    \hline
  \end{tabular}
  \caption{Potential goodput improvements with one Interest.}
  \label{fig:potential-improvement}
\end{figure}

\subsection{Link-layer Interest bundling avoids design complexities}
\label{subsec:avoid-complexity}
\begin{figure}[!ht]
  \vspace{1mm}
  \centering
  \includegraphics[width=0.95\columnwidth]{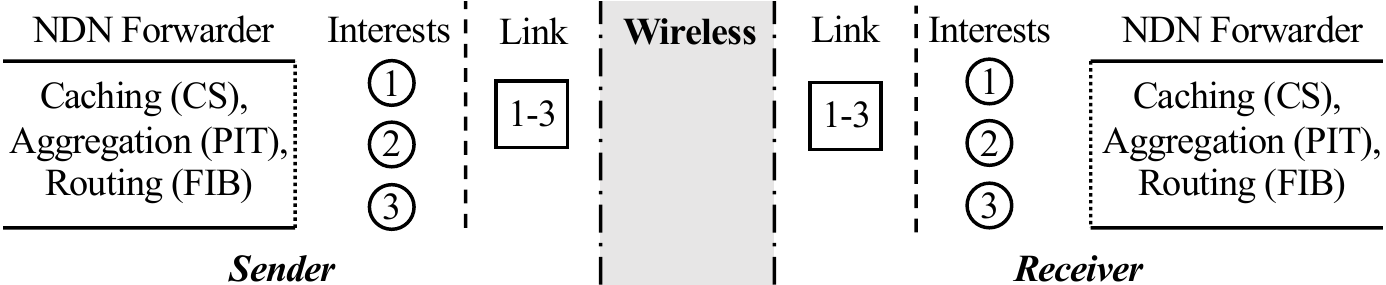}
  \caption{Example of masking link-layer changes.}
  \label{fig:ll-masking}
\end{figure}

A link-layer-based solution for Interest bundling in NDN can \textit{mask} any wireless channel usage optimization to the upper layers. In Fig.~\ref{fig:ll-masking}, the sender-side (left-half) forwarder performs cache matching, aggregation at the Pending Interest Table (PIT) and routing with the Forwarding Information Base (FIB). After that, three Interest packets with sequence numbers 1, 2, 3 are ready to be sent out. The link-layer bundles them into one and sends to the wireless network card. The receiver-side (right-half) link-layer unbundles the three Interest packets such that they are identical to the sender-side ones before bundling and passes them to the forwarder, which then repeats the steps similar to the sender-side. These steps preserve the content store, Interest aggregation behavior, and routing operations. Thus the forwarder is unaware of any link-layer changes.

Using this approach in BLEnD should allow for almost seamless integration between different forwarding techniques, as well as different link-layer encoding-decoding variations.

\section{Related Works}
\label{sec:related-works}

The primary incentive of BLEnD comes from recent works in TCP-IP such as \cite{tack-Tong} which implements a periodic ACK mechanism to increase the Data flow's channel utilization. A similar adaptation in NDN is challenging given that the Data flow here is opposite to TCP-IP. Moreover, an end-to-end congestion control would break the hop-by-hop design philosophy of NDN. \cite{Rahman-dil-manet} shows the downsides of channel contention in ad-hoc networks. It also motivates the need to reduce Interest channel occupancy to avoid a forced limitation on the congestion window increment, limiting the goodput.

Recent works by \cite{Ding-video} and \cite{Fan-infocom-video} propose similar ideas of using one Interest to retrieve a sequence of consecutive Data packets. They reduce the number of Interest packets sent in the wireless channel for retrieving the same amount of Data packets. However, neither specifies details of ensuring NDN's hop-by-hop flow balancing, i.e., one Interest at the forwarding plane fetches one Data packet at each link; how to deal with congestion control and reliability.
\section{The Link-layer Design for Interest Bundling}
\label{sec:design}

BLEnD's link-layer encoding-decoding over a single hop consists of two segments, a) sender-side encoding and b) receiver-side decoding. Notations used throughout the rest of the paper are in Table~\ref{tab:notations}. We have implemented the design in the GenericLinkService of the \texttt{face} subsystem in NFD~\cite{afanasyev2014nfd}.

\begin{table}[!t]
  \caption{Notations}
  \scriptsize
  \label{tab:notations}
  \centering
  \begin{tabular}{l|l}
    \hline
    \textbf{Notation} & \textbf{Meaning}\\
    \hline
    $cwnd$ & Congestion window \\
    \hline
    $BI$ & Bundle Interval \\
    \hline
    \texttt{bTAG} & 64 bit Bundle TAG \\
    \hline
    RTx & Retransmission \\
    \hline
    $seq$ & Interest Sequence number \\
    \hline
    $LSS$ & Last start sequence \\
    \hline
    $LES$ & Last end sequence \\
    \hline
    $SS$ & Start sequence \\
    \hline
    $ES$ & End sequence \\
    \hline
    $PSS$ & Previous start sequence \\
    \hline
    $PES$ & Previous end sequence \\
    \hline
    $DSS$ & Decode start sequence \\
    \hline
    $DES$ & Decode end sequence \\
    \hline
  \end{tabular}
\end{table}

\subsection{Sender-side encoding}
\label{subsec:sender-encoding}
\begin{figure}[!t]
  \vspace{1mm}
  \centering
  \includegraphics[width=0.98\columnwidth]{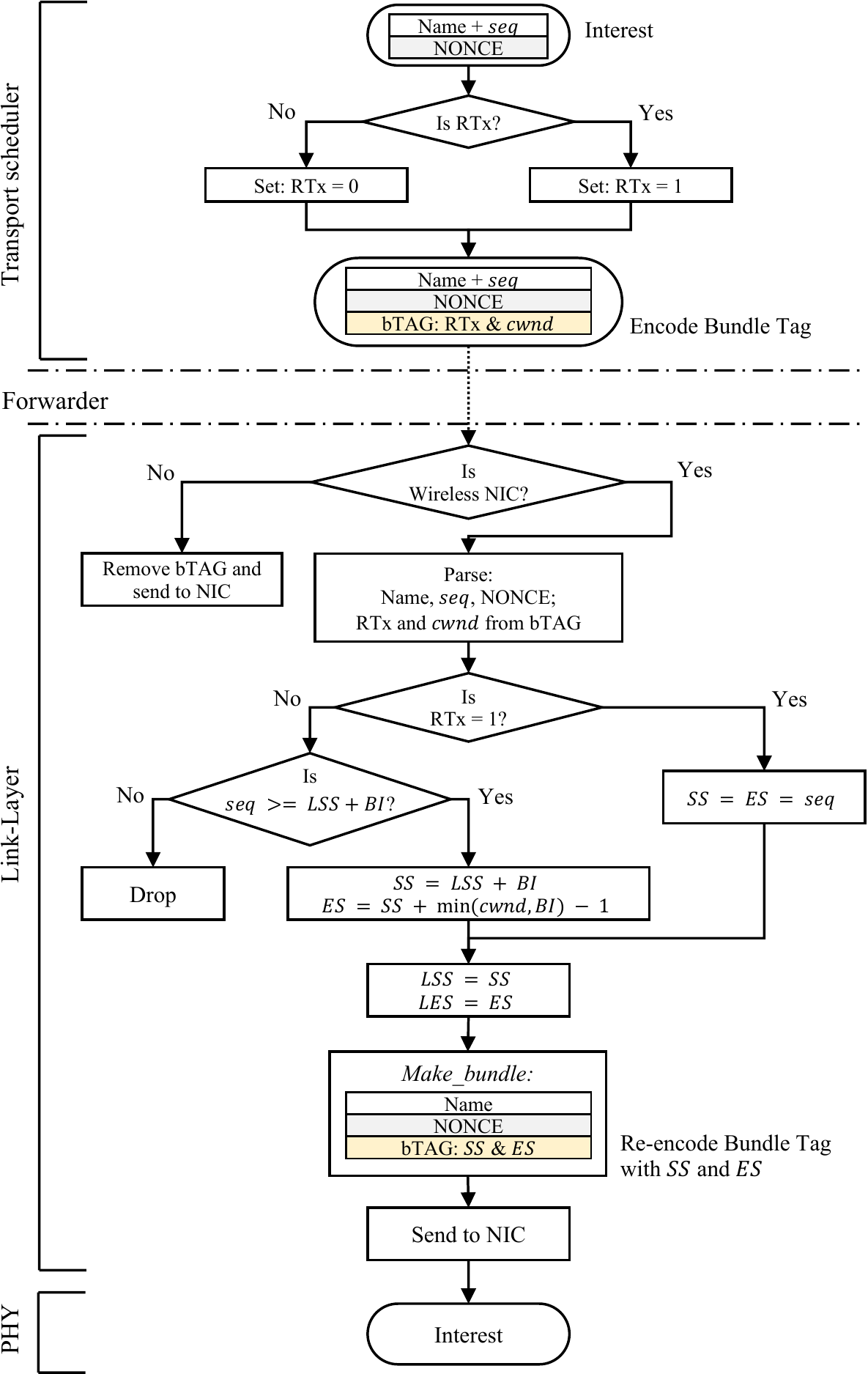}
  \caption{Sender side link-layer Interest bundling (encode) pipeline.}
  \label{fig:sender-encode-flowchart}
\end{figure}

The encoder logic flowchart is in Fig.~\ref{fig:sender-encode-flowchart}. The transport scheduler is responsible for scheduling Interests received from the application based on a congestion control algorithm while the link-layer receives a packet from the network forwarder and sends it out to the NIC. The \texttt{Name} of an Interest carries a $seq$ suffix, representing a chunk's sequence number of a larger Data file. The link-layer must know two information to bundle an Interest: an RTx or not and the current $cwnd$. The transport can easily pass this information to the link-layer. Thus, this cross-layer process has two sub-steps, 1) Interest tagging at the transport and b) link-layer encoding. 

\begin{figure}[!t]
  \vspace{1mm}
  \centering
  \includegraphics[width=0.98\columnwidth]{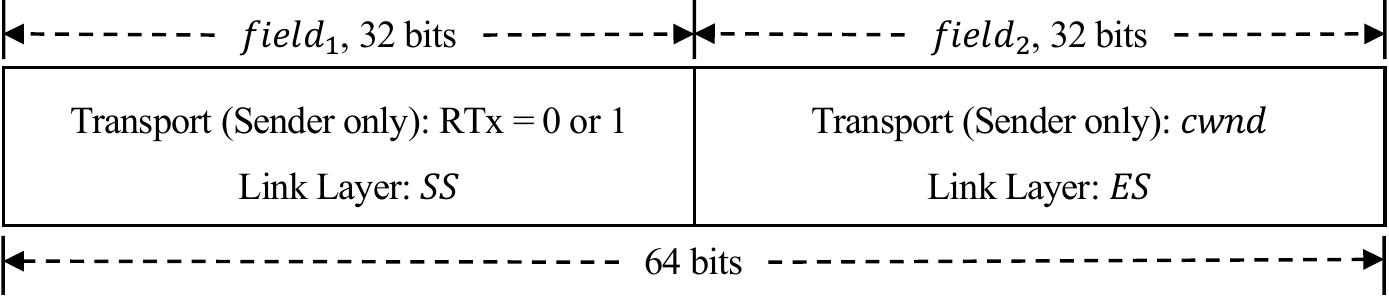}
  \caption{Breakdown of 64 bit bundle tag (bTAG).}
  \label{fig:bundle-tag}
\end{figure}

\subsubsection{Interest tagging at the transport}
\label{subsubsection:transport-tagging}
The transport scheduler passes the RTx and $cwnd$ information by adding a 64-bit tag to an Interest called the \texttt{BundleTag} or \texttt{bTAG}. We defined \texttt{bTAG} in the \texttt{ndn-cxx} library following the \texttt{TLV} format in NDN Packet Specifications~\cite{NDN-packet}. The upper 32 bits ($field_1$) represent the RTx in the form of a boolean flag (first bit here is set to 1 if yes, otherwise 0), while the lower 32 bits carry the $cwnd$ ($field_2$). The \texttt{bTAG} structure is in Fig.~\ref{fig:bundle-tag}.

When the transport has an Interest to schedule, it marks the RTx field accordingly, updates the $cwnd$ filed, adds the \texttt{bTAG} to the Interest, and forwards to the network-layer forwarder. The forwarder performs its usual routines~\cite{Yi-stateful} and currently has no rule for handling \texttt{bTAG}. Thus, the network layer continues to carry the \texttt{bTAG} to the link layer.

\subsubsection{Link-layer encoding}
\label{subsubsection:link-encoding}

Upon receiving an Interest from the network forwarder, the link-layer should first detect the NIC type (wired or wireless) and continue encoding logic if it is wireless. Otherwise, it removes the \texttt{bTAG} and sends it to the NIC like a regular Interest packet. We do this because BLEnD is intended for wireless links, and almost all wired NICs are full-duplex. If the NIC is wireless, the link-layer continues the encoding process, parses the \texttt{Name}, $seq$, \texttt{NONCE} fields, and decodes the \texttt{RTx} and $cwnd$ from the \texttt{bTAG}.

The $LSS$, $LES$, and $BI$ are initialized per \texttt{Name} prefixes and represent encoding states of each complete Data file (e.g., /yt/ml.mp4) without the sequence numbers. We call this link-layer state maintenance table the Encoding Information Table or EIT, and each entry looks like a tuple, 
\begin{equation*}
  <Name, BI, LSS, LES>.
\end{equation*}

Now, if the link-layer sees an Interest as an \texttt{RTx} ($filed_1=1$), it sets the start ($SS$) and end ($ES$) sequences for the intended bundle to $seq$. Doing so ensures that an RTx packet will be always sent out for fast recovery from a possible loss event. If the condition fails, the link-layer will check if $seq \ge LSS+BI$. It is a crucial part of the encoder as it uses a fixed bundle-interval ($BI$) to encode a single Interest that represents a set of \textit{future} $BI-1$ Interest packets, plus the current one. The check also helps block and drop any future Interest with a $seq$ already included in the past encoded packet.

Next, if a $seq$ qualifies as a new bundle candidate, the link-layer sets $SS=LSS+BI$ and updates $ES$ such that, $ES-SS+1=min(cwnd, BI)$. It ensures that when we are at the last bundle, the transport assigned the $cwnd$ to be the remaining packets and possibly lower than $BI$. We then update the $LSS$ and $LES$ for the next potential bundle.

In the next step, the link-layer encodes the new bundle with the parsed \texttt{Name} and \texttt{NONCE} while also keeping the lifetime of the Interest intact. It then re-encodes the \texttt{bTAG} by setting $field_{1}=SS$ and $field_2=ES$. The encoding is now complete, and the link-layer sends it to the NIC, which goes out as a regular Interest packet to the wireless medium.

\subsection{Receiver-side decoding}
\label{subsection:receiver-decoding}

\begin{figure}[!t]
  \vspace{1mm}
  \centering
  \includegraphics[width=0.98\columnwidth]{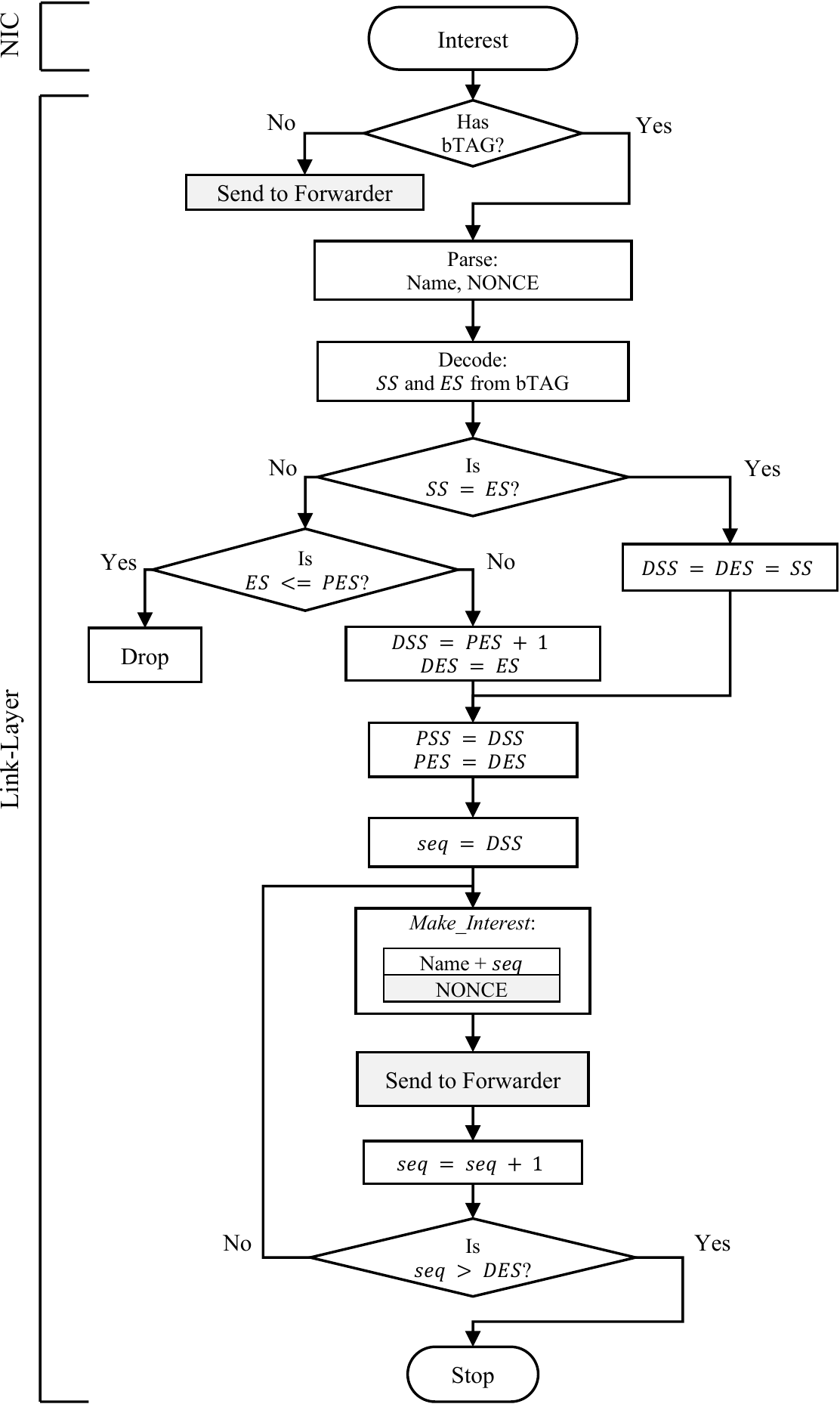}
  \caption{Receiver side link-layer bundled-Interest decoding pipeline.}
  \label{fig:receiver-decode-flowchart}
\end{figure}

The receiver-side decoder logic flowchart is in Fig.~\ref{fig:receiver-decode-flowchart}. Upon receiving a new Interest from the NIC, the link-layer first checks if it has the \texttt{bTAG} or not. If not, directly sends to the network forwarded and starts decoding process otherwise. $PSS$ and $PES$ are the start and end sequences of a previous bundle, respectively, mapped to per \texttt{Name} tuple at the Decoding Information Table, or DIT,
\begin{equation*}
  <Name, PSS, PES>.
\end{equation*}
It is similar to the EIT at the encoder, except for the $BI$.

The decoder then parses the \texttt{Name}, \texttt{NONCE}, $SS$, and $ES$. If $SS=ES$ (an RTx), it sets the decoding start ($DSS$) and end ($DES$) sequences to $SS$. If $SS\ne ES$, the decoder checks if $ES<=PES$; if true, it drops the bundle (previously decoded). Otherwise, it further processes by updating $DSS$ and $DES$ to one more than $PES$ and $ES$, respectively. 

\begin{figure}[!t]
  \vspace{1mm}
  \centering
  \includegraphics[width=0.95\columnwidth]{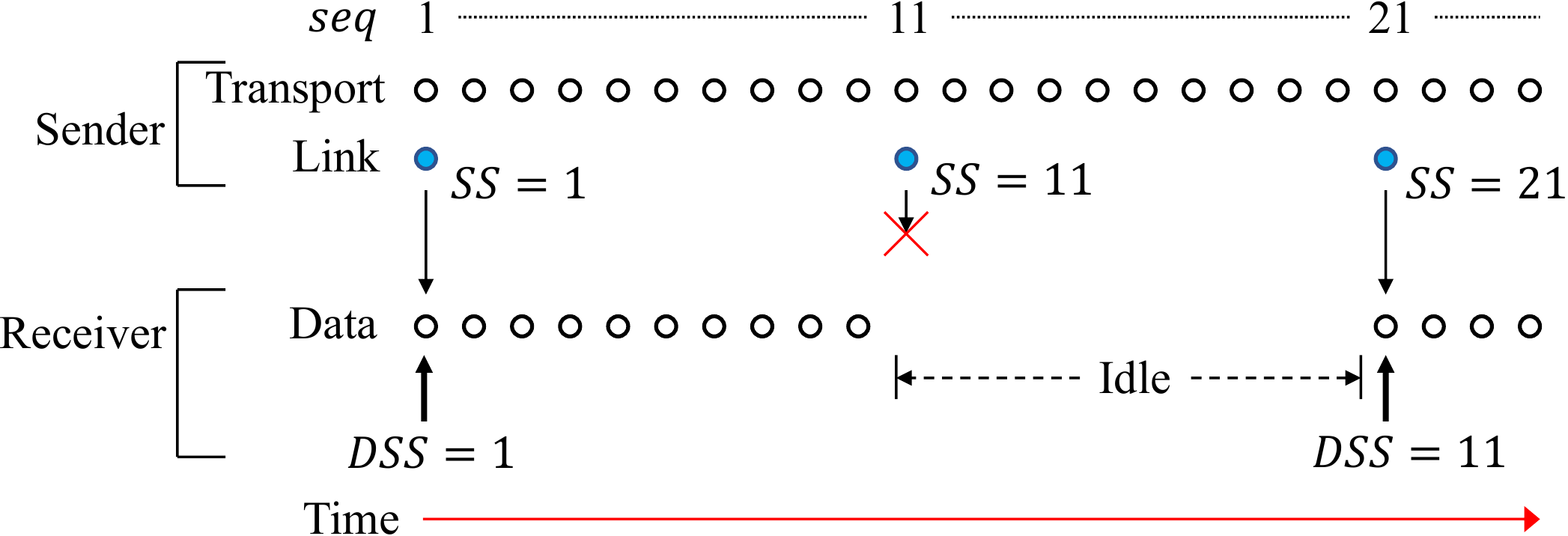}
  \caption{Potential receiver-side idle period on bundled-Interest loss and gap filling. $BI=10$ at the sender.}
  \label{fig:decode-gap}
\end{figure}

Here, $DSS=PES+1$ is a crucial recovery technique for a potential previously lost bundled-Interest. To explain this, we show an example in Fig.~\ref{fig:decode-gap} where the rows represent the packet sequence numbers, the sender's, and the receiver's current states over time (from left to right, without showing RTx). If $BI=10$, there are three potential bundles with $SS$ = 1, 11, and 21, respectively, in the solid-blue circles. However, if the second bundle ($SS=11$) is lost, the receiver will stay idle (currently, $PES=10$). Only when the next bundle with $SS=21$ comes in, the receiver will resume data sending, and to recover from the gap in Data packets and minimize Interest RTx, it sets $DSS=PES+1=11$.

Following the recovery mechanism, the decoder updates $PSS$ and $PES$. At this stage, the decoder is ready to construct new Interest packet(s) from the parsed and decoded information of the bundled Interest. It runs a loop by initializing $seq=DSS$ and repeating while $seq<=DES$. Each iteration constructs a new Interest by adding the $seq$ as a suffix to the \texttt{Name}, adds the \text{NONCE} and preserves the lifetime (not shown). It then forwards the newly formed Interest to the network forwarder and increments $seq$ by one. When the loop breaks, the decoding process is complete, and the link-layer is ready to decode a new bundled-Interest.

One noticeable change from regular Interest packets is that a bundled Interest carries only one \texttt{NONCE} (first Interest of a bundle). \texttt{NONCE} helps detect Interest loops, and BLEnD's design leads to the same \texttt{NONCE} at the PIT of the receiver for the different sequences. However, in infrastructure Wi-Fi, a consumer will not get loops, and nodes beyond the access point will carry the \texttt{NONCE} from the consumer's link-generated bundle, preserving loop detection. We leave BLEnD in ad-hoc networks for future work.

%
\section{Performance Analysis}
\label{sec:evaluation}

\subsection{Simulation setup}
\label{subsec:simulation}
We conducted simulations with BLEnD by using the stationary two-nodes setup in Sec.~\ref{subsec:interest-reduction} with infrastructure Wi-Fi. It enables frame-level acknowledgments on unicast transmissions. Thus, we used a data-centric self-learning strategy~\cite{Rahman-daf-manet} as our network-layer forwarding. We ran our simulations to download a 100 MB file using the AIMD and CUBIC congestion controllers included with ndnSIM (v2.7). We collected results for the default (or no bundle) Interest-Data exchange and our BLEnD implementation.

\subsection{Effect of wireless link capacity}
\label{subsec:link-capacity}

\begin{figure}[!t]
  \vspace{1mm}
  \centering
  \includegraphics[width=0.82\columnwidth]{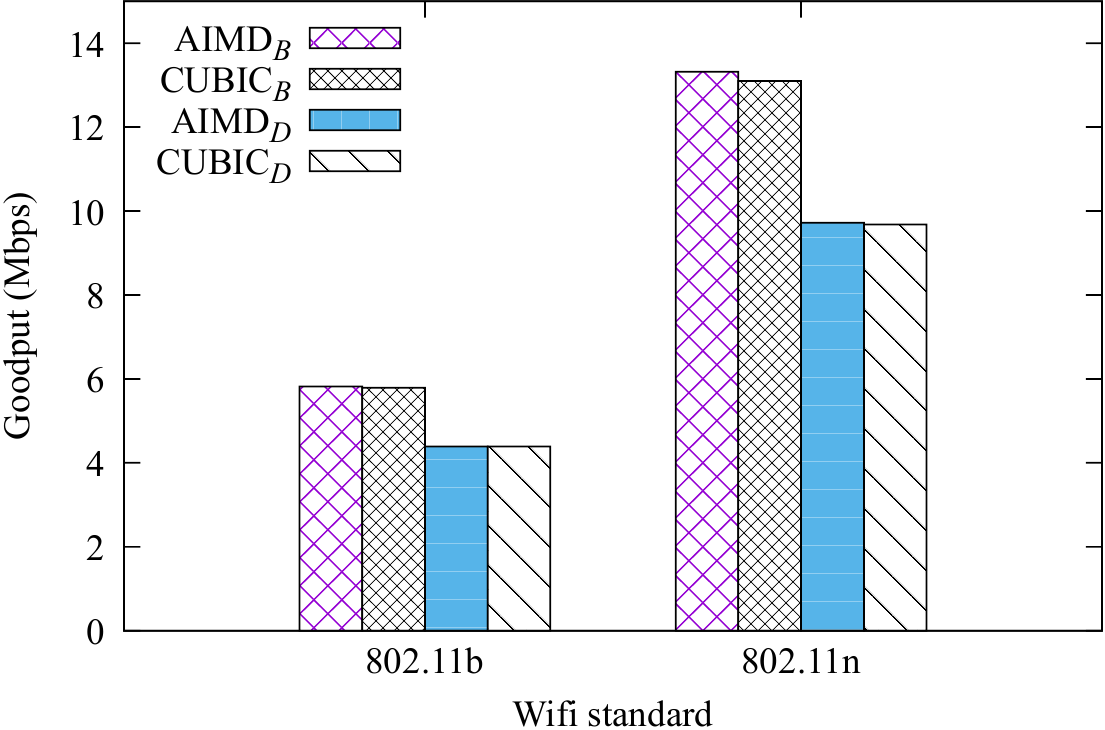}
  \caption{Simulated measured goodput with different IEEE 802.11 standards and PHY capacity (802.11b set at 11 Mbps and 802.11n set at 24 Mbps). Subscript shows B = bundled-interest and D = default or no bundling. $BI=15$.}
  \label{fig:wifi-goodput}
\end{figure}

Fig.~\ref{fig:wifi-goodput} shows a goodput comparison between BLEnD and default Interest-Data exchange. We see that with AIMD, BLEnD achieves 32.57\% and 37.03\% more goodput in 802.11b and 802.11n, respectively, than the default design. CUBIC with BLEnD also achieves 31.9\% and 35.33\% more goodput in 802.11b and 802.11n, respectively, which is very close to AIMD. CUBIC shows a slightly lower goodput than AIMD in BLEnD because of its more conservative window adaptation approach. Moreover, ~\cite{Rahman-dil-manet} shows an initial window increase spike in wireless networks for NDN. It leads to more loss in AIMD than CUBIC in default exchange, but CUBIC's conservative approach balances their overall performance. Moreover, bundling Interests lowers the Interest loss from the initial window spike, therefore yielding better goodput.

The results show that even with a simple design like BLEnD, we can achieve more than 30\% goodput improvement. Furthermore, BLEnD's goodput with AIMD here is roughly 20\% and 41\% lower than the one Interest result we saw in Fig.~\ref{fig:potential-improvement}. Such results are expected as we have only presented the initial design. The results prove that further improvements or optimizations should achieve performance close to the dummy application goodput.

\subsection{Effect of bundle size}
\label{subsec:bundle-size}
\begin{figure}[!t]
  \centering
  \subfloat[Goodput.]{
    \includegraphics[width=0.464\columnwidth]{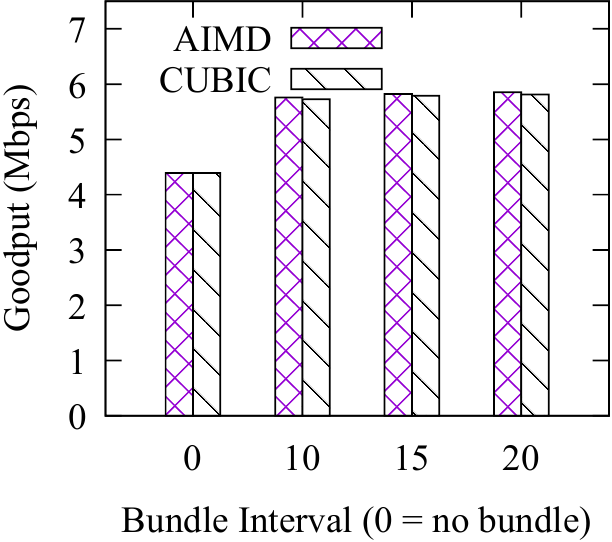}
    \label{fig:bundle-goodput}
  }
  \hfill
  \subfloat[Link-layer Tx-events.]{
    \includegraphics[width=0.47\columnwidth]{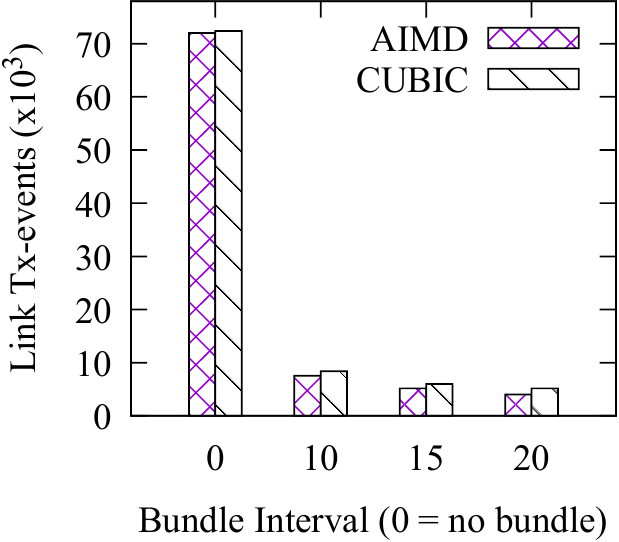}
    \label{fig:bundle-sent}
  }
  \caption{Effect of Bundle Interval (BI) on goodput and sender's link-layer packet transmissions (Tx) using 802.11b (11 Mbps). BI=0 represents no Interest bundling. Link-layer Tx includes application or transport RTx packets.}
  \label{fig:bundle-size-effect}
\end{figure}

We also analyzed the effect of different bundle-Interval or $BI$ values at the sender side encoder on the goodput and link-layer transmission (Tx) events. Currently, we are using a fixed size $BI$, and the results are in Fig.~\ref{fig:bundle-size-effect}.

Fig.~\ref{fig:bundle-goodput} shows that an increasing $BI$ value yields a small goodput gain but using a small $BI$ can significantly improve goodput compared to no bundling. It is expected and further proves that freeing up channel usage by reducing signal packet frequency allows more Data packets to occupy it and, in turn, improve the application throughput.

Fig.~\ref{fig:bundle-sent} further proves that increasing bundle size contributes to a fair amount of reduction of the Tx events at the sender. In AIMD, increasing $BI$ from 10 to 15 decreases Tx events by 31.4\%. We also see higher Tx in CUBIC than in AIMD with BLEnD, which results from a conservative window adaptation approach combined with a critical premature timeout effect, which we further discuss in the next section (\ref{subsec:retx-timeout}).

\section{Challenges and Future Works}
\label{sec:challenges-future-works}

Although our early BLEnD design already improves goodput by up to 37\% compared to the default NDN approach, some critical cases need attention to improve performance and avoid unexpected breakdowns. We listed a few here for future research and present some potential benefits of using BLEnD.

\subsection{Premature retransmission timeout}
\label{subsec:retx-timeout}
Continuing from Sec.\ref{subsec:bundle-size}, we observed fairly high RTx events at the sender's link. In infrastructure mode or ad-hoc mode with RTS/CTS enabled, the wireless NIC waits for the channel to free up. Thus, some data packets queue up at the receiver NIC as the network layer of the Data node (in our case, the access point) processes unbundled Interest packets almost immediately. As a result, over a single hop, heavy RTT fluctuations happen at the consumer node. However, continuously incoming data packets keep the RTO low, and consequently, many Interest packets timeout prematurely, leading to more RTx events. Higher premature timeout events may lead to more RTx and duplicate Data, causing BLEnD to perform poorly.

Although our primary focus was to design a link-layer encoder-decoder and not play around with transport or application, we ran some simulations with different multipliers for the RTT's variance parameter to calculate round-trip timeout or RTO. The testing results with different RTT variance multipliers (we call it $\gamma$) are in Fig.~\ref{fig:rttvm}.

\begin{figure}[!ht]
  \scriptsize
  \centering
    \begin{tabular}{c|c|c|c|c|c|c}
      \hline
      & \textbf{$\gamma=$} & RTx & \textbf{$App_{sent}$} & \textbf{$P_{sent}$} & \textbf{$C_{rcv}$} & \textbf{Goodput (Mbps)} \\
      \hline
      \hline
      AIMD & 4 & 629 & 72450 & 72450 & 71847 & 13.32 \\
          & 10 & 590 & 72411 & 72411 & 71821 & 13.33 \\
          & 20 & 580 & 72401 & 72401 & 71821 & 13.34 \\
        
      \hline
      \hline
      CUBIC & 4 & 2567 & 74388 & 74388 & 72315 & 13.1 \\
            & 10 & 1982 & 73803 & 73803 & 71996 & 13.2 \\
            & 20 & 1616 & 73437 & 73437 & 71827 & 13.25 \\
      \hline
    \end{tabular}
  \caption{Effect of RTT variance multiplier ($\gamma$) using 802.11n, file=100MB, BI=15, $Data_{pkts}$=71821. $\gamma=4$ is default.}
  \label{fig:rttvm}
\end{figure}

Here $App_{sent}$ is the complete application or transport issued Interests including RTx, $P_{sent}$ is the access point or producer Tx events, and $C_{rcv}$ is the amount of data packets received by the consumer's link (including duplicates). $Data_{pkts}$ is the number of data packets or chunks that the consumer application expects (for 100 MB file, 1460 bytes of payload per packet). We see that $\gamma=4$ (default RTO multiplier) leads to a relatively high number of RTx events, leading to many redundant Data Tx and increasing channel contention. As a result, many of them are lost at the PHY-layer ($P_{sent}-C_{rcv}$).

With $\gamma=10$, we can see a drop in RTx, $App_{sent}$, $P_{sent}$, and $C_{rcv}$, proving our claim of the premature timeout events. Increasing $\gamma$ to 20 in CUBIC further decreases RTx and, thus, Tx events go down as well. CUBIC reduces RTx by 22.7\% and 37\% with $\gamma=10$ and $\gamma=20$, respectively, compared to using $\gamma=4$. For AIMD, $\gamma=20$ leads to negligible or no change of send and receive events. It is expected as CUBIC has higher RTx (308\% more with $\gamma=4$) and Tx events for its conservative approach than AIMD. With $\gamma>4$, Goodput slightly goes up as Tx and wireless loss go hand-to-hand.

To avoid the premature RTO effect, we believe a rate-based congestion control approach would be a better fit for using BLEnD. The transport can then \textit{space out} each scheduling event so that RTT sampling start-time is late enough to reduce excessive RTT fluctuations over time.

\subsection{Bundle loss recovery and dynamic interval}
\label{subsec:gap-fill}
In Sec.\ref{subsection:receiver-decoding}, Fig.~\ref{fig:decode-gap}, we showed how the receiver fills missing sequences of a lost bundle packet. Unfortunately, it does not minimize nor remove the idle period. To solve this, we think a rate-based approach at the sender's link-layer encoder can be adopted to sample the \textit{inter-Data gap} over time \cite{Amadeo-SIRC}. Thus, instead of always blocking Interests with $seq \ge LSS+BI$, the encoder can estimate a bundle loss earlier than the decoder node and send out a new bundle when it detects an inter-Data gap sample is significantly higher than an estimated moving average over time. It should also help dynamically adjust $BI$ as the inter-Data gap can be different in other network topologies, and wireless link capacities.

\subsection{Sequenced data files and a robust \texttt{bTAG}}
\label{subsec:sequence-files}
Next, BLEnD currently works on files with same base-name (e.g., \texttt{/yt/bi.mp4}) with different data chunks represented by increasing sequence numbers (e.g., \texttt{/yt/bi.mp4/1}, \texttt{/yt/bi.mp4/2} and so on.) However, applications may not use such sequencing altogether. We expect further research to improve BLEnD for supporting a wide variety of applications. Moreover, the current \texttt{bTAG} is limited to 64 bits. Using a dynamic length \texttt{bTAG} should be able to include segmented bundles with multiple RTx $seq$ numbers. We leave the study for future work.

\subsection{Potential to improve performance in large networks}
\label{subsec:multi-hop}
Channel contention in multi-hop or multi-node wireless tactical networks is more severe than single hop. Even with a congestion window limit~\cite{Rahman-dil-manet}, Interest frames have a similar channel acquisition overhead of Data frames. 

\begin{figure}[!ht]
  \vspace{1mm}
  \centering
  \includegraphics[width=0.90\columnwidth]{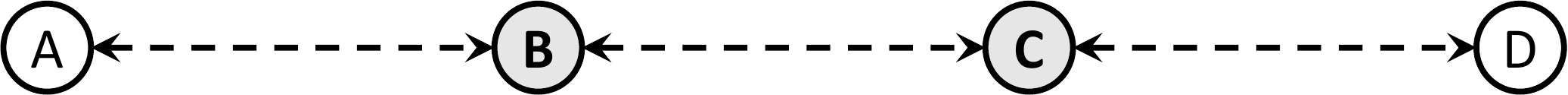}
  \caption{Example of a multi-node wireless network.}
  \label{fig:multi-node}
\end{figure}

Fig.~\ref{fig:multi-node} shows four wireless nodes with the dotted lines as their wireless links. Nodes $B$ and $C$ sends Interests to nodes $A$ and $D$, respectively. Without Interest bundling, $C$ and $B$ become each others' potential \textit{hidden-stations}. As a result, packet collision can occur when $A$ and $D$ transmit data packets, while $C$ and $B$ transmit Interest packets, respectively. Bundling Interests can significantly reduce the hidden-station problem and improve throughput. Thus, BLEnD can scale well beyond a single-hop, in multi-hop wireless networks, and under mobility where collision and contention are major bottlenecks. We leave multi-hop or multi-node wireless experiments for future works.

\subsection{Hop-by-hop transport}
\label{subsec:hop-transport}
BLEnD also asks for further research in hop-by-hop transport management where each hop can independently bundle Interests (possibly over multi-path for load-balancing) and improve network performance even more. Hop-by-hop flow/congestion control should further improve reliability and minimize loss in mission critical networks.

\subsection{Hardware testing}
\label{subsec:hardware}
Finally, we look forward to testing BLEnD on real hardware with effects on signal quality, link capacity, power consumption to understand the bundling behavior in a better way and analyze the challenges to formulate appropriate optimizations.

%
\section{Conclusion}
\label{sec:conclusion}

Using one-Interest one-Data philosophy in wireless networks leads to downgraded application performance in NDN because of shared channel capacity. Thus, bundling multiple Interest packets into one to fetch multiple Data packets can extensively improve goodput. On the other hand, implementing an Interest bundling technique either at the application, transport, or the network forwarder can completely break the hop-by-hop philosophy of NDN's stateful forwarding plane. Our novel BLEnD design tries to solve both problems by encoding-decoding bundled Interests at the link layer. The design is simple enough, requiring straightforward changes to the NFD while leaving the forwarding pipeline untouched. Preliminary simulation results show promising improvements over default link-layer design and pave a path for numerous research challenges in the future.


\bibliographystyle{IEEEtran}  
\bibliography{refs}

\begin{thebibliography}{10}
\providecommand{\url}[1]{#1}
\csname url@samestyle\endcsname
\providecommand{\newblock}{\relax}
\providecommand{\bibinfo}[2]{#2}
\providecommand{\BIBentrySTDinterwordspacing}{\spaceskip=0pt\relax}
\providecommand{\BIBentryALTinterwordstretchfactor}{4}
\providecommand{\BIBentryALTinterwordspacing}{\spaceskip=\fontdimen2\font plus
\BIBentryALTinterwordstretchfactor\fontdimen3\font minus
  \fontdimen4\font\relax}
\providecommand{\BIBforeignlanguage}[2]{{%
\expandafter\ifx\csname l@#1\endcsname\relax
\typeout{** WARNING: IEEEtran.bst: No hyphenation pattern has been}%
\typeout{** loaded for the language `#1'. Using the pattern for}%
\typeout{** the default language instead.}%
\else
\language=\csname l@#1\endcsname
\fi
#2}}
\providecommand{\BIBdecl}{\relax}
\BIBdecl

\bibitem{goldsmith2005wireless}
A.~Goldsmith, \emph{Wireless communications}.\hskip 1em plus 0.5em minus
  0.4em\relax Cambridge university press, 2005.

\bibitem{ndn-ccr}
L.~Zhang, A.~Afanasyev, J.~Burke, V.~Jacobson, kc~claffy, P.~Crowley,
  C.~Papadopoulos, L.~Wang, and B.~Zhang, ``{Named Data Networking},''
  \emph{ACM Computer Communication Reviews}, Jun. 2014.

\bibitem{ieee-mac}
\BIBentryALTinterwordspacing
``Wireless lan medium access control (mac) and physical layer (phy)
  specifications,'' \emph{IEEE Standards Association.}, 2016. [Online].
  Available: \url{https://ieeexplore.ieee.org/document/7786995}
\BIBentrySTDinterwordspacing

\bibitem{tack-Tong}
T.~Li, K.~Zheng, K.~Xu, R.~A. Jadhav, T.~Xiong, K.~Winstein, and K.~Tan,
  ``Tack: Improving wireless transport performance by taming acknowledgments,''
  ser. SIGCOMM '20.\hskip 1em plus 0.5em minus 0.4em\relax New York, NY, USA:
  ACM, 2020, p. 15–30.

\bibitem{altman2003novel}
E.~Altman and T.~Jim{\'e}nez, ``Novel delayed ack techniques for improving tcp
  performance in multihop wireless networks,'' in \emph{IFIP PWC}.\hskip 1em
  plus 0.5em minus 0.4em\relax Springer, 2003, pp. 237--250.

\bibitem{floyd-profile}
S.~Floyd and E.~Kohler, ``{Profile for Datagram Congestion Control Protocol
  (DCCP) Congestion Control ID 2: TCP-like Congestion Control},'' RFC 4341,
  Mar. 2006.

\bibitem{landstrom2007}
S.~Landstr\"{o}m and L.-r. Larzon, ``Reducing the tcp acknowledgment
  frequency,'' \emph{SIGCOMM Comput. Commun. Rev.}, vol.~37, no.~3, p. 5–16,
  Jul. 2007.

\bibitem{nadh2013improving}
K.~L. Nadh, Y.~S. Krishna, and K.~N. Rao, ``Improving tcp performance with
  delayed acknowledgments over wireless networks: a receiver side solution,''
  in \emph{ARTCom 2013}.\hskip 1em plus 0.5em minus 0.4em\relax IET, 2013, pp.
  195--201.

\bibitem{Rahman-daf-manet}
\BIBentryALTinterwordspacing
M.~A. Rahman and B.~Zhang, ``On data-centric forwarding in mobile ad-hoc
  networks: Baseline design and simulation analysis,'' 2021. [Online].
  Available: \url{https://arxiv.org/abs/2105.07584}
\BIBentrySTDinterwordspacing

\bibitem{Yi-stateful}
C.~Yi, A.~Afanasyev, I.~Moiseenko, L.~Wang, B.~Zhang, and L.~Zhang, ``A case
  for stateful forwarding plane,'' \emph{Comput. Commun.}, vol.~36, no.~7, pp.
  779--791, Apr. 2013.

\bibitem{Ding-video}
X.~Ding, W.~Yang, and F.~Wu, ``Adaptive video streaming transmission mechanism
  based on wireless ndn,'' in \emph{HPCC/SmartCity/DSS 2020}, 2020, pp.
  944--949.

\bibitem{Fan-infocom-video}
F.~Wu, W.~Yang, J.~Ren, F.~Lyu, X.~Ding, and Y.~Zhang, ``Adaptive video
  streaming using dynamic ndn multicast in wlan,'' in \emph{IEEE INFOCOM WKSHPS
  2020}, 2020, pp. 97--102.

\bibitem{NDN-packet}
\BIBentryALTinterwordspacing
``Ndn packet format specification.'' [Online]. Available:
  \url{https://named-data.net/doc/NDN-packet-spec/current/}
\BIBentrySTDinterwordspacing

\bibitem{mastorakis2017ndnsim}
S.~Mastorakis, A.~Afanasyev, and L.~Zhang, ``On the evolution of {ndnSIM}: an
  open-source simulator for {NDN} experimentation,'' \emph{ACM Computer
  Communication Review}, Jul. 2017.

\bibitem{afanasyev2014nfd}
\BIBentryALTinterwordspacing
A.~Afanasyev, J.~Shi, B.~Zhang, L.~Zhang, I.~Moiseenko, Y.~Yu, W.~Shang
  \emph{et~al.}, ``Nfd developer's guide.'' [Online]. Available:
  \url{https://named-data.net/publications/techreports/ndn-0021-11-nfd-guide/}
\BIBentrySTDinterwordspacing

\bibitem{Rahman-dil-manet}
M.~A. Rahman and B.~Zhang, ``On the analysis of adaptive-rate applications in
  data-centric wireless ad-hoc networks,'' in \emph{2021 IEEE 46th Conference
  on Local Computer Networks (LCN)}, 2021, pp. 503--510.

\bibitem{Amadeo-SIRC}
M.~Amadeo, A.~Molinaro, C.~Campolo, M.~Sifalakis, and C.~Tschudin, ``Transport
  layer design for named data wireless networking,'' in \emph{2014 IEEE INFOCOM
  WKSHPS}, 2014, pp. 464--469.

\end{thebibliography}

\end{document}